\documentclass[%
 reprint,
 amsmath,amssymb,
 aps,
]{revtex4-1}

\usepackage[utf8]{inputenc}

\usepackage{graphicx}% Include figure files
\usepackage{dcolumn}% Align table columns on decimal point
\usepackage{bm}% bold math
\begin{document}

\preprint{APS/123-QED}

\title{Multiple scattering production of lepton-pairs \\in ultraperipheral heavy-ion collisions}% Force line breaks with \\

\author{Sevgi Karadağ}
 \altaffiliation{karadags@itu.edu.tr}%Lines break automatically or can be forced with \\

\author{Mehmet Cem Güçlü}%
 \email{guclu@itu.edu.tr}
\affiliation{%
 Department of Physics,Istanbul Technical University\\
}%

\date{\today}% It is always \today, today,

\begin{abstract}
Ultra-relativistic peripheral collisions of heavy ions at LHC can produce copies of numbers of lepton pairs via the two-photon process. Since the energies of the heavy ions are so high, multi-pair
production cross-sections of the light leptons especially electrons are quite large so that it is possible to measure them experimentally. To calculate the multi-pair production probabilities,
first, we should have an impact parameter dependence cross-section. We have obtained a well-behaved impact parameter dependence cross-section and by using the Monte-Carlo methods, we have calculated multi-pair production cross-sections of electrons and muons in Pb-Pb heavy-ion collisions at LHC energies. We have also used some experimental restrictions in our calculation to compare our findings with the experimental results.
%\begin{description}
%\end{description}
\end{abstract}

\maketitle

%\tableofcontents

\section{\label{sec:level1}Introduction}

One of the main goals of the Relativistic Heavy-Ion Collider (RHIC) and Large Hadron Collider (LHC) is to investigate the quark-gluon plasma which is the confined state of the partonic matter. In central collisions of the heavy ions, dileptons are produced as a result of the quark-antiquark interaction (Drell-Yan process), and these dileptons interact only with electromagnetically with the particles \cite{2,11}. The lepton-hadron interaction is very small hence they carry direct information from the first formation of space-time. 

The ultra-peripheral collisions of heavy-ions are described by an impact parameter $b$ greater than the sum of the radius of the colliding nuclei. When two ultra-relativistic heavy ions pass near each other, the intense electromagnetic fields are very strong so that they pull lepton pairs from the vacuum. Lepton pairs coming from the quark-gluon plasma can be mixed with the electromagnetically produced lepton pairs. Therefore, it is extremely important to investigate the electromagnetic production of lepton pairs in detail.

From the pure science point of view, relativistic heavy-ion collisions provide an opportunity to study non-perturbative quantum electrodynamics (QED). In such collisions, colliding energies and electric charges are so high that the coherent electromagnetic production of lepton pairs using heavy ions is fundamentally different from other production mechanisms using light particles at high energies. In the heavy-ion case, the coupling constant is strongly enhanced due to the large charge. Nonetheless, the problem of reliable lowest order perturbative calculations for the lepton pair production has been used as input into design models for the RHIC and LHC.

We start with the quantum-field theory, and derive a classical field method based on the Feynman perturbation theory. In these calculations, we should be careful that the momentum transfer of the photons is much smaller than the momentum carried by the nuclei. This assumption is valid for the coherent production of lepton pairs through two-photon processes in relativistic heavy-ion collisions. By using these lowest-order diagrams, coupling lepton fields to classical electromagnetic fields have been evaluated exactly with the help of the Monte Carlo method and analytical techniques \cite{bs, bb, mcg, mcgjl}, and have been predicted cross-sections consistent with the experiments for the lepton pair production.

On the other hand, by using the lowest-order perturbation theory to the production of electron-positron pairs with heavy ions at high energies and small impact parameters results in probabilities and cross-sections which violate the unitarity \cite{2,18}. It is therefore clear that low-order perturbative calculations alone are not adequate for smaller impact parameters at the RHIC and especially at the LHC energies. Therefore, higher-order terms must be included in the calculations. 
 
In literature, electron-positron pair production from the vacuum is treated by applying the equivalent-photon approximation (EPA). In 1924, Fermi replaced the time-dependent electric field by an approximately equivalent spectrum of on-shell photons. The problem of the moving Coulomb field of the charged particles was done in the 1930s. Weizsacker and Williams modified this problem for the relativistic particles. In this approximation, the equivalent-photon flux created by the relativistic charged particle is obtained via a Fourier decomposition of the electromagnetic interaction \cite{18, ww}. Cross-sections are calculated by folding the elementary, real two-photon cross-section for the pair-production process using the equivalent-photon flux produced by each ion. The pair-production cross-section is relatively easier to compute using the equivalent-photon approximation than by calculating the two-photon diagram. When we compare these two methods, the results for the total cross-section are reasonably accurate provided the incident-particle Lorentz factor $\gamma$ is much greater than one and the energy transferred via the photon is much less than $\gamma$. However, the details of differential cross-sections, spectra, and impact-parameter dependence differ, especially when complicated numerical cuts in the coordinates are applied in order to compare with experiments. Among the shortcomings of this approximation is an undetermined parameter which corresponds to the minimum impact parameter or the maximum momentum transfer, which makes it difficult to get specific results. As such, the method loses the applicability at impact parameters less than the Compton wavelength of the lepton\cite{ahtb, bhatk}, which is the region of greatest interest for the study of non-perturbative effects. Reference\cite{baur} reviews the application of the equivalent-photon method in relativistic heavy-ion collisions.

The recent experiments at RHIC and LHC provided us several data about electromagnetic production of lepton pairs from the ultra-peripheral collisions of the heavy-ions. The rapidity, invariant mass, and transverse momentum distributions are measured with the experimental cut-offs. In our calculations, we have estimated these observables with the experimental limits and compared them with the recent ALICE and ATLAS data for the electron-positron and muon-antimuon pair production \cite{ATLAS_mu, ALICE_el, JadamALICE, MAatlas, MD, JadamSTAR, LAstar, 
SAphenix}. We have also predicted the electromagnetic production of double $e^{+}e^{-}$  and $\mu^{+}\mu^{-}$ pair productions.

\section{Formalism}
\begin{figure}
	\centering
	\includegraphics[scale=0.5]{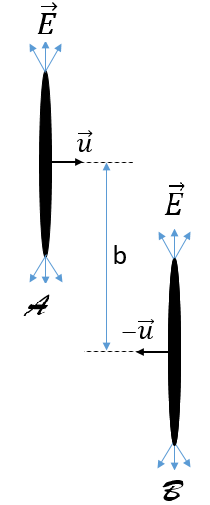}
\caption{Collision of two heavy-ions at relativistic velocities}
\label{two-heavy-ions}
\end{figure}
The cross sections for producing a final state $X$ in Equivalent-Photon Method is written as the integral lover the "photon fluxes" of each of the ions times
the on-shell two-photon to the $X$ cross section,
\begin{equation}
\sigma=\int \int d\omega_{1} d\omega_{2} n_{A_{1}}(\omega_{1}) n_{A_{2}}(\omega_{2}) \sigma_{\gamma \gamma \rightarrow X}(\omega_{1} \omega_{2})
\end{equation}
where $n(\omega)$ is the equivalent photon distribution, and $A$ is the colliding nuclei. The EPA is an approximation to the two-photon external field model that we studied in this article.

Fig.\ref{two-heavy-ions} shows two heavy ions moving with relativistic velocities $u$ along the $z-$axis in opposite directions to each other. They undergo the Lorentz contraction due to their high relativistic velocities. The distance between the radii of the nuclei is shown by the impact parameter $b$. Throughout this paper, we use the natural units with $\hbar=c=m=e=1$.

\begin{figure*} 
\centering
\includegraphics[scale=0.5]{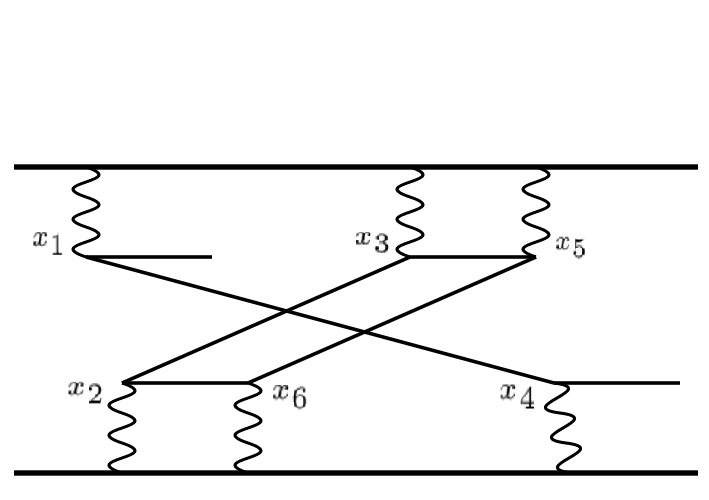} \qquad
\includegraphics[scale=0.5]{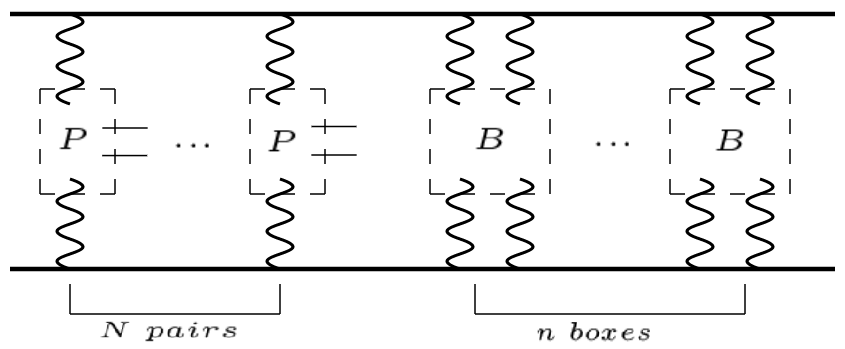} \qquad
\begin{minipage}{1.0 \textwidth}
\centering 
\caption{Figure(a) contains one pair creation and one box. All of the diagrams in figure(b) which we include are in the form of $N$ pair creations and $n$ boxes.}
\label{feynmans}
\end{minipage}
\end{figure*}

Since the physical processes such as lepton-pair production or annihilation occur with the interaction of fields, the lepton pair production from the electromagnetic field is expressed by the interaction lagrangian density which depends on the classical $4$-vector potential $A^{\mu}$:
\begin{equation} \label{1}
\mathcal{L}_{int}(x)=- \bar{\psi}(x) \gamma_{\mu} \psi(x) A^{\mu}(x)
\end{equation}
where $A^\mu$ can be written separately for two colliding nuclei $\mathcal{A}$ and $\mathcal{B}$ 
\begin{equation} \label{2}
A^{\mu}(q,b)=A^{\mu}_{\mathcal{A}}(q,b)+A^{\mu}_{\mathcal{B}}(q,b).
\end{equation}
4-vector potential in momentum space can be written as,
\begin{align} \label{3,4,5}
A^0_{(\mathcal{A},\mathcal{B})}(q) = & -8 \pi^2  \gamma^2 (Ze)  G_E(q^2) f_z(q^2)  \nonumber \\ & \times \exp \bigg(\pm i \mathbf{q_{\perp}} \cdot \frac{\mathbf{b}}{2} \bigg) \frac{\delta(q_0 \mp \beta q_z)}{q_z^2+\gamma_{\perp}^2} , \\ A^z_{(\mathcal{A},\mathcal{B})}(q)=&\pm \beta A^0_{(\mathcal{A},\mathcal{B})}(q), \\ \mathbf{A}^{\perp}_{(\mathcal{A},\mathcal{B})}(q)=&0
\end{align}
where $f_z(q^2)$ and $G_E(q^2)$ are the form factors. Form factors are crucial for the cross-section calculation especially for heavy lepton-pair production in electromagnetic collisions. 
In our calculations, we have used the Fermi (Woods-Saxon) distribution for the charge distribution of the nucleus,
\begin{equation} \label{6}
\rho(r)=\dfrac{\rho_0}{1+\exp \big( \frac{r-R}{a}\big)}
\end{equation}\\
where $R$ is the radius of the nucleus, and $a$ is a quantity related to the thickness of the nucleus shell \cite{Sengul}. Also, $\rho_0$ can be calculated by normalization. 

Using the expressions given so far, the cross-section depending on the impact parameter $b$ for the lepton pair production can be written 
\begin{equation} \label{8}
\sigma=\int d^2 \mathbf{b} \sum_{k>0} \, \sum_{q<0} \, \big|\langle \chi_k^{(+)} | S| \chi_q^{(-)} \rangle \big|^2
\end{equation}
where the summation over the states $k$ is restricted to those above
the Dirac sea, and the summation over the states $q$ is restricted to
those occupied in the Dirac sea.

The result for $S_{\mathcal{A} \mathcal{B}}$ is
\begin{align} \label{D14}
\langle \chi_k^{(+)} | S_{\mathcal{A} \mathcal{B}}| & \chi_q^{(-)} \rangle = \nonumber \\  &\dfrac{i}{2 \beta} \int \dfrac{d^2 \mathbf{p}_{\perp}}{(2 \pi)^2} \exp \bigg[i \bigg( \mathbf{p}_{\perp}- \dfrac{(\mathbf{k}_{\perp}+\mathbf{q}_{\perp})}{2}  \bigg) \cdot \mathbf{b} \bigg]  \nonumber \\ & \times F(\mathbf{k}_{\perp}-\mathbf{p}_{\perp}; \omega_{\mathcal{A}}) \, F(\mathbf{p}_{\perp}-\mathbf{q}_{\perp}; \omega_{\mathcal{B}}) \, \mathcal{T}_{kq}(\mathbf{p}_{\perp}; \beta)
\end{align}

where, in above equation the functions $F( \bf q, \omega)$ and $\mathcal{T}_{kq}(\mathbf{p}_{\perp}; \beta)$ can be written explicitly as

\begin{equation} \label{D15}
F(\mathbf{k}_{\perp}-\mathbf{p}_{\perp}; \omega_{\mathcal{A}})= \dfrac{4 \pi Z \gamma^2 \beta^2}{\omega_{\mathcal{A}}^2+\gamma^2 \beta^2(\mathbf{k}_{\perp}-\mathbf{p}_{\perp})^2}
\end{equation}

\begin{equation} \label{D16}
F(\mathbf{p}_{\perp}-\mathbf{q}_{\perp}; \omega_{\mathcal{A}})= \dfrac{4 \pi Z \gamma^2 \beta^2}{\omega_{\mathcal{B}}^2+\gamma^2 \beta^2(\mathbf{p}_{\perp}-\mathbf{q}_{\perp})^2}
\end{equation}

\begin{align} \label{D17}
\mathcal{T}_{kq} & (\mathbf{p}_{\perp}; \beta) = \nonumber \\ &i \sum_s \sum_{\sigma_p} \bigg[(E_p^{(s)}- \bigg( \dfrac{E_k^{(+)}+E_q^{(-)}}{2} \bigg) + \beta \bigg( \dfrac{k_z-p_z}{2} \bigg) \bigg]^{-1} \nonumber \\ & \times \langle u_{\sigma_k}^{(+)}|(1-\beta \alpha_z)|u_{\sigma_p}^{(s)} \rangle \, \langle u_{\sigma_p}^{(s)}|(1+\beta \alpha_z)|u_{\sigma_q}^{(-)} \rangle.
\end{align}
Here, $ |\chi_k^{(+)} \rangle $ refers to the positive-energy spinors, and $|\chi_q^{(-)}\rangle$ refers to the negative-energy spinors. $S$ is the scattering matrix and expressed as a series expansion. In our previous works, we have obtained exact cross-section expressions and tried to calculate it with Monte-Carlo integration methods \cite{semi-analytic,erkan}.
However, the Bessel function in this equation is rapidly oscillating, especially for large impact-parameters, it makes difficult to calculate the cross-section numerically.
On the other hand,  impact-parameter dependent cross-section is very helpful for understanding the physical phenomena occurring in heavy-ion collisions.
Therefore, to overcome this problem the cross-section expression can be divided into two parts.
\begin{equation} \label{14}
\dfrac{d\sigma}{d b}= \int_0^{\infty} qdq \, b J_0(qb) \, \mathcal{F}(q)
\end{equation}
Here, $\mathcal{F}(q)$ is a 9-dimensional integral (Eq.\eqref{15}) that can be calculated with Monte-Carlo integration method \cite{semi-analytic}. 

When this expression is calculated for different values of $q$ with the Monte-Carlo integration method, we can fit the results to a smooth function for $\mathcal{F}(q)$. \\
\begin{widetext}
\begin{align} \label{15} 
\mathcal{F}(q)=&\frac{\pi}{8 \beta^2} \sum_{\sigma_k} \sum_{\sigma_q} \int_0^{2 \pi} d \phi_q \int \frac{dk_z dq_z d^2 k_{\perp} d^2 K d^2 Q}{(2 \pi)^{10}} \bigg\{ F \Big( \frac{\mathbf{Q}-  \mathbf{q}}{2}; \omega_{\mathcal{A}} \Big) F \big(-\mathbf{K};\omega_{\mathcal{B}} \big) \mathcal{T}_{kq} \Big(\mathbf{k}_{\perp}-\frac{\mathbf{Q}-\mathbf{q}}{2}; \beta \Big) \nonumber \\&+F \Big( \frac{\mathbf{Q}-  \mathbf{q}}{2}; \omega_{\mathcal{A}} \Big) F \big(-\mathbf{K};\omega_{\mathcal{B}} \big) \mathcal{T}_{kq} \big(\mathbf{k}_{\perp}- \mathbf{K}; -\beta \big) \bigg\} \bigg\{F \bigg( \frac{\mathbf{Q} + \mathbf{q}}{2}; \omega_{\mathcal{A}} \bigg) F \big(-\mathbf{q}- \mathbf{K};\omega_{\mathcal{B}} \big) \mathcal{T}_{kq} \bigg(\mathbf{k}_{\perp}-\frac{\mathbf{Q}+ \mathbf{q}}{2}; \beta \bigg) \nonumber \\ & +F \Big(\frac{\mathbf{Q}+  \mathbf{q}}{2}; \omega_{\mathcal{A}} \Big) F \big(-\mathbf{q}- \mathbf{K};\omega_{\mathcal{B}} \big) \mathcal{T}_{kq} \big(\mathbf{k}_{\perp}+\mathbf{q}- \mathbf{K}; -\beta \big) \bigg\} 
\end{align}
\end{widetext}
Here, $\omega_{\mathcal{A}}$ and $\omega_{\mathcal{B}}$ are the frequencies of the virtual photons: 
\begin{subequations} \label{D13}
\begin{align}
\omega_{\mathcal{A}} &=\dfrac{E_q^{(-)}-E_k^{(+)}+\beta(q_z-k_z)}{2} \\ \omega_{\mathcal{B}} &=\dfrac{E_q^{(-)}-E_k^{(+)}-\beta(q_z-k_z)}{2}.  
\end{align}
\end{subequations}

The transformations used in this equation are
\begin{subequations}
\begin{align} \label{g13}
\mathbf{Q}&=2\mathbf{k}_{\perp}-\mathbf{p}_{\perp}-\mathbf{p}_{\perp}^{\prime} \\ \mathbf{q}&=\mathbf{p}_{\perp}-\mathbf{p}_{\perp}^{\prime} \\ \mathbf{K}&=-\mathbf{p}_{\perp}+\mathbf{q}_{\perp}
\end{align}
\end{subequations}

\begin{figure*} 
\centering
	\includegraphics[scale=0.5]{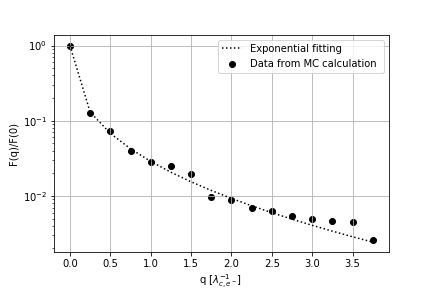} \quad
	\includegraphics[scale=0.5]{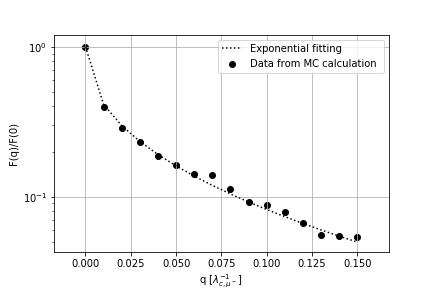}
	\begin{minipage}{1.0 \textwidth}
	\caption{The change of $\mathcal{F}(q)/\mathcal{F}(0)$ depended on $q$ variable for both (a)$e^{+}e^{-}$  and (b)$\mu^{+}\mu^{-}$  pairs production from $Pb-Pb$ collision at LHC energies(Here, $\lambda_c$ represents the compton wavelengths of electron and muon).}
	\label{a_electron-muon5TeV}
	\end{minipage}
\end{figure*}

We have done all these calculations for different energies for producing electron and muon pairs, and have shown the results in Fig. \ref{a_electron-muon5TeV}. In our previous works, 
these smooth fits were generally in the form of $\sim e^{-a \, q}$. However, when we improved our calculations, we have found that the most appropriate fits should be in the form of
\begin{equation} \label{16}
\mathcal{F}(q)=\mathcal{F}(0)e^{-a_{1} \,q^{a_{2}}}
\end{equation} 
where, $\mathcal{F}(0)$ gives the total cross-section $\sigma_T$ at $q=0$. The parameters
$a_{1}$ and $a_{2}$ are obtained from the suitable fit functions. 

To calculate the differential cross-section, we can insert the fit function obtained in Eq.\eqref{16} into Eq.\eqref{14}
\begin{equation} \label{17}
\dfrac{d\sigma}{d b}= \mathcal{F}(0) \int_0^{\infty} qdq \, b J_0(qb) \, e^{-a_{1} \,q^{a_{2}}}.
\end{equation}
By using the differential cross-section, one can calculate the total cross-section, multi-pair production probabilities, and the other observables.

\section{Multi-pair creation}
There are important reasons to understand multiple-pair production. When we increase the colliding energies, one-pair production cross sections become very large, and multi-pair
production probability has also an important contribution to the observed cross-sections. For high charged ions such as $Au$, $Pb$,  and for very high energies such as LHC, the lowest order perturbative calculations are not sufficient to calculate the probabilities of lepton pairs creation. Therefore, we need higher-order Feynman diagrams to evaluate the cross-sections.
However, it is very cumbersome to get expressions and to evaluate them. On the other hand, we can make some approximations and obtain sufficient results. 
In literature and in our previous works, details of the approximations are shown in a clear manner \cite{bb, mcgjl, ahtb}.
As an approximation, only the Fermion loops which have exactly two vertices attached to each of the ions are included. 
As an example, in Fig. \ref{feynmans}{a}, one pair creation and one "box" are shown.
To calculate the N pair creations, we should include N pair creations and n boxes as shown in  Fig. \ref{feynmans}{b} . We did not include the additional interactions such as the final state interactions of a created lepton with one of the ions. Therefore, when we calculate the sum of all this diagrams, we obtain
the Poisson form for the multi-pair production:

\begin{equation} \label{19}
P_N(b)=\dfrac{\mathcal{P}(b)^N e^{-\mathcal{P}(b)}}{N!}
\end{equation}
where $N$ is the number of pairs, and $\mathcal{P}(b)$ is the probability of producing a single pair from the second-order Feynman diagrams: 
\begin{equation} \label{18}
\mathcal{P}(b)=\dfrac{1}{2 \pi b} \dfrac{d \sigma}{db}
\end{equation}
As we discussed above, this probability is not accurate for large values of charges and energies of the colliding ions. However, by obtaining the Poisson form 
for the multi-pair production, the perturbative result for the production of a single pair is modified by the exponential factor.
In order to obtain $N$ pair cross-section, we just need to integrate the $N$-pair probability $P_N(b)$
 \citep{mcg, ahtb, semi-analytic, erkan},

\begin{equation} \label{20}
\sigma_{N_{pair}}=\int d^2 b \, P_N(b).
\end{equation}

\begin{table} 
\centering
\caption{$\sigma_T$ values for the production $e^{+}e^{-}$ and $\mu^{+}\mu^{-}$ pairs from $Pb-Pb$ collision at LHC energies. The first group shows the single lepton-pair production, while the second group shows double lepton-pair production. The last group is for triple pair production.}
\scalebox{0.85}{%
\begin{tabular}{l l l}
\hline 
\hline
\rule{0pt}{2.5ex} Lepton pairs \quad\qquad\qquad\qquad\qquad & $2.76$ TeV \qquad\qquad\qquad\qquad &\ $5.02$ TeV \\ & (barn) \qquad\qquad\qquad\qquad & (barn)\\[0.5ex] 
\hline 
\rule{0pt}{2.5ex} $e^{+}e^{-}$  &  $1.67 \times 10^5$\qquad\qquad\qquad\quad & $2.17 \times 10^5$ \\ [0.5ex]
\rule{0pt}{2.5ex} $\mu^{+}\mu^{-}$ & $1.67$ \qquad\qquad\qquad\qquad & $2.30$   \\ [0.5ex]
\\ 
\rule{0pt}{2.5ex} $e^{\pm}$-$e^{\pm}$ & $3.29 \times 10^3$ \qquad\qquad\qquad\qquad & $4.26 \times 10^3$   \\ [0.5ex] 
\rule{0pt}{2.5ex} $\mu^{\pm}$-$\mu^{\pm}$ & $1.72 \times 10^{-3}$ \quad\qquad\qquad\qquad &  $2.22 \times 10^{-3}$ \\ [0.5ex] 
\rule{0pt}{2.5ex} $e^{\pm}$-$\mu^{\pm}$ & $1.61 \times 10^{-1}$ \qquad\qquad\qquad\qquad &   $2.07 \times 10^{-1}$   \\[1.5ex]
\\
\rule{0pt}{2.5ex} $e^{\pm}$-$e^{\pm}$-$e^{\pm}$ & $5.76 \times 10^2$ \qquad\qquad\qquad\qquad & $7.42 \times 10^2$   \\ [0.5ex]
\rule{0pt}{2.5ex} $\mu^{\pm}$-$\mu^{\pm}$-$\mu^{\pm}$ & $7.73 \times 10^{-6}$\quad\qquad\qquad\qquad &  $11.7 \times 10^{-6}$ \\ [0.5ex]
\hline
\hline
\end{tabular}}
\label{sigmaT-values}
\end{table}

\begin{table} 
\centering
\caption{The ratio between $\sigma_T$ values of single, double and triple pairs production from $Pb-Pb$ collision. The first group shows electron pairs while the second group shows muon pairs.}
\scalebox{0.85}{%
\begin{tabular}{l l r}
\hline
\hline 
\rule{0pt}{2.5ex} Lepton pairs \quad\quad\qquad\qquad\qquad\qquad & $2.76$ TeV \qquad\qquad\qquad\qquad & $5.02$ TeV \\[0.5ex] 
\hline 
\rule{0pt}{2.5ex} $\sigma_{1p}^{e^{-}} / \sigma_{2p}^{e^{-}} $  &  $0.51 \times 10^2$\qquad\qquad\qquad\quad & $0.51 \times 10^2$ \\ [0.5ex]
\rule{0pt}{2.5ex} $\sigma_{1p}^{e^{-}} / \sigma_{3p}^{e^{-}} $  &  $0.29 \times 10^3$\qquad\qquad\qquad\quad & $0.29 \times 10^3$ \\ [0.5ex]
\rule{0pt}{2.5ex} $\sigma_{2p}^{e^{-}} / \sigma_{3p}^{e^{-}} $  &  $0.57 \times 10^1$\qquad\qquad\qquad\quad & $0.57 \times 10^1$ \\ [0.5ex]
\\
\rule{0pt}{2.5ex} $\sigma_{1p}^{\mu^{-}} / \sigma_{2p}^{\mu^{-}} $  &  $0.97 \times 10^3$\qquad\qquad\qquad\quad & $1.04 \times 10^3$ \\ [0.5ex]
\rule{0pt}{2.5ex} $\sigma_{1p}^{\mu^{-}} / \sigma_{3p}^{\mu^{-}} $  &  $ 0.22 \times 10^6 $\qquad\qquad\qquad\quad & $0.20 \times 10^6$ \\ [0.5ex]
\rule{0pt}{2.5ex} $\sigma_{2p}^{\mu^{-}} / \sigma_{3p}^{\mu^{-}} $  &  $0.22 \times 10^3$\qquad\qquad\qquad\quad & $0.19 \times 10^3$ \\ [0.5ex]
\hline
\hline
\end{tabular}}
\label{sigmaT-values-a-ratio}
\end{table}

\section{Results and discussions}
In two photon-external field model that we study here, the most important task is to evaluate the Eq. \eqref{15} numerically. We have done this by using
Monte-Carlo integration method. We have used sufficiently large random numbers to evaluate it accurately.

\begin{figure*}
	\centering
	\includegraphics[scale=0.5]{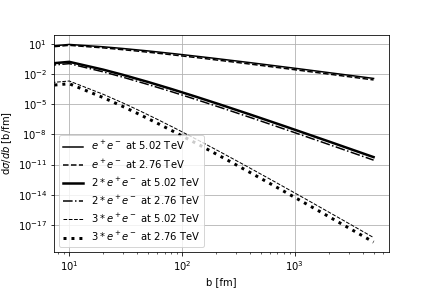}\quad
	\includegraphics[scale=0.5]{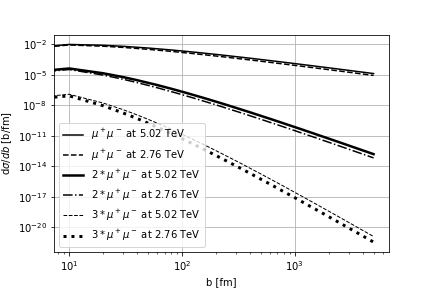}	
	\caption{The change of the cross-section of (a)electron and (b)muon for single, double and triple pairs production from $Pb-Pb$ collision dependent on the impact parameter $b$ at LHC energies, $\sqrt{s_{NN}}=2.76$ and $5.02$ TeV.}
\label{single-double-triple-el-mu-2.76-5.02}
\end{figure*}

\begin{figure*} 
\centering
	\includegraphics[scale=0.5]{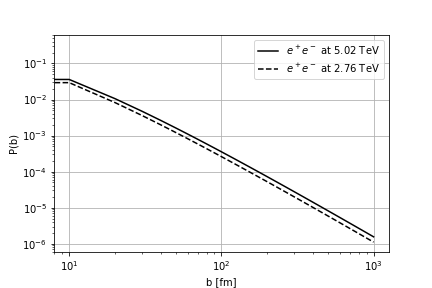} \quad
	\includegraphics[scale=0.5]{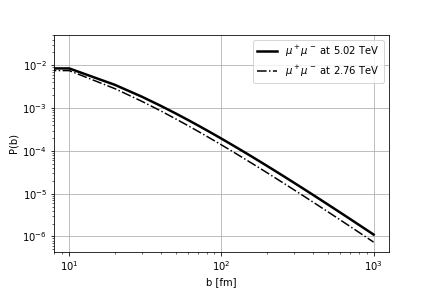}
	\begin{minipage}{0.85 \textwidth}
	\caption{The probabilities for (a)$e^{+}e^{-}$  and (b)$\mu^{+}\mu^{-}$ pairs production from $Pb-Pb$ collision at LHC energies.}
	\label{ol_electron-muon}
	\end{minipage}	
\end{figure*}

Figs. \ref{a_electron-muon5TeV}{a} and \ref{a_electron-muon5TeV}{b} show the values of $\mathcal{F}(q)/\mathcal{F}(0)$ for each values of $q$  in Pb-Pb collisions at LHC for electron and muon pairs productions, respectively. A smooth function is then fit to the calculated numbers. By using this fit function, we can calculate the impact parameter dependent cross-sections and probabilities of the producing lepton pairs. These calculations are shown in Figs.\ref{single-double-triple-el-mu-2.76-5.02}{a} and \ref{single-double-triple-el-mu-2.76-5.02}{b} for electron and muon 
pairs, respectively. In these plots, the colliding energies of the $Pb$ ions are considered as $\sqrt{s_{NN}}=2.76$ and  $5.02$ TeV and we have plotted the single pair, double pairs, and triple pairs differential cross-sections. After obtaining the differential cross sections, from the Eqs. \eqref{19} and \eqref{18}, we can calculate the probabilities of $N$ pair productions.
These probabilities are plotted in Figs. \ref{ol_electron-muon}{a} and \ref{ol_electron-muon}{b} as a function of impact parameter for above energies.

The calculated values of total cross-sections $\sigma_T$ for single, double and triple lepton-pairs production are shown in Table \ref{sigmaT-values}. It is clear that, when
the colliding energies increase, total cross sections increase as a result of this. In our calculations, we have used the Fermi (Woods-Saxon) distribution for the charge distribution of the nucleus,  Eq.\eqref{6}. Since the Compton wavelength of the muon is comparable with the size of the nucleus, form factors are very crucial in the calculations.
For simplicity, we have tabulated the ratio between the multi-pair total cross-section values in Table \ref{sigmaT-values-a-ratio}. Although, electron single-pair production is the dominant
process, double and triple pairs production mechanism has also significant contribution to the total pair production process. On the other hand, due to the higher mass,
the gaps between the ratios of muon multi-pair cross sections are higher compare to electron-pair productions.

\begin{table}[b] 
%\centering
\caption{First line shows the experimental and calculated results of total cross-sections at the rapidity range $|y|<2.4$ for $\mu^{+}\mu^{-}$ pair at $\sqrt{s_{NN}}=5.02$ TeV(ATLAS). The other lines show the results for $e^{+} e^{-}$ pair at the rapidity range $|y|<0.9$ at $\sqrt{s_{NN}}=2.76$ TeV(ALICE). }
\scalebox{0.83}{%
\begin{tabular}{l l l r}
\hline
\hline
\rule{0pt}{2.5ex} Lepton pair \quad & Our result  \quad & LHC \cite{ATLAS_mu,ALICE_el} \quad \qquad \qquad & STARLIGHT \cite{STARLIGHT} \\ & ($\mu b$) \qquad\qquad & ($\mu b$)\qquad\qquad & ($\mu b$)\\[0.5ex]
\hline
\rule{0pt}{2.5ex} $\mu^{+} \mu^{-} $ & $ 25 $  & $32.2 \pm  0.3(stat)^{+4.0}_{-3.4}(sys)$ & $ 31.64 \pm 0.04(stat)$\\ [1.0ex]
\\
\rule{0pt}{2.5ex} $e^{+} e^{-} $ & $ 117 $  & $91 \pm 10(stat)^{+11}_{-8}(sys)$ & $ 77$\\ $(3.7<W<10)$ \\[0.0ex]
\\
\rule{0pt}{2.5ex} $e^{+} e^{-} $ & $ 103 $  & $154 \pm 11(stat)^{+17}_{-11}(sys)$ & $ 128$ \\ $(2.2<W<2.6)$\\ [0.1ex]
\hline
\hline
\end{tabular}}
\label{ATLAS-ALICE-total-cross}
\end{table}

So far, we have not used any experimental cuts in our calculations on parameters, such as rapidity, transverse momentum and invariant mass of the produced leptons.
However, due to some technical difficulties, experimentally,  cross-sections are measured for limited values of the parameters. Therefore, in order to compare
our calculations with the experimental results, we have also restricted our calculations within the experimental limits.

\begin{figure*}
	\centering
	\includegraphics[scale=0.5] {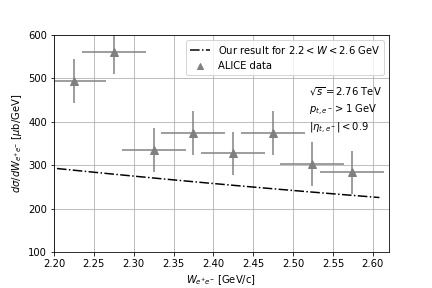} \quad
	\includegraphics[scale=0.5] {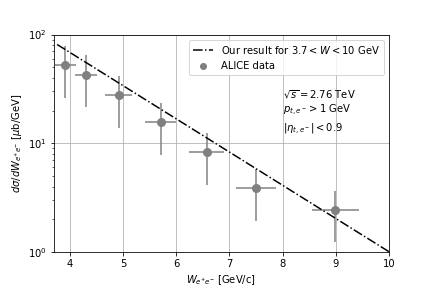}
\caption{The (a)low and (b)high invariant-mass distribution of $e^{+}e^{-}$ pair in $Pb-Pb$ heavy-ion collision at LHC ($\sqrt{s_{NN}}=2.76$ TeV) energy together with the ALICE experiment data\cite{ALICE_el}.}
\label{cut_e_inv_1380}
\end{figure*}

In ALICE experiment, the kinematic restrictions for the measured $e^{+} e^{-}$ pair production from Pb-Pb collision at $\sqrt{s_{NN}}=2.76$ TeV are $|y|<0.9$ for rapidity of all electrons and pairs, $2.2>W_{e^+ e^-}>10$ GeV for invariant mass, and $p_{t,{e^-}}>1$ GeV for transverse momentum. We have applied these restrictions to our calculations, and shown our results in Figs.\ref{cut_e_inv_1380}{a} and \ref{cut_e_inv_1380}{b} for both low invariant-mass range $2.2<W_{e^+ e^-}<2.6$ GeV and high invariant-mass range $3.7<W_{e^+ e^-}<10$ GeV, respectively. Then, we integrated these curves over invariant-mass ranges to find total cross-section values, and showed the results in Table \ref{ATLAS-ALICE-total-cross}. While our result slightly overestimate  the experimental result for the high invariant-mass range, we obtained a lower value for the low invariant-mass range. On the other hand, STARLIGHT calculations \cite{STARLIGHT} slightly lower than the experimental results.

\begin{figure}
	\centering
	\includegraphics[scale=0.5] {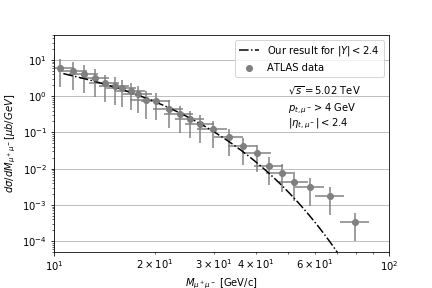}
\caption{The invariant mass distribution of $\mu^{+}\mu^{-}$ pair in $Pb-Pb$ heavy-ion collision at LHC ($\sqrt{s_{NN}}=5.02$ TeV) energy together with the ATLAS experiment data \cite{ATLAS_mu}.}
\label{cut_mu_2510}
\end{figure}

In ATLAS experiment, the kinematic restrictions for the production of $\mu^{+}\mu^{-}$ pair in Pb-Pb collision at $\sqrt{s_{NN}}=5.02 $ TeV  are $|y|<2.4$ for rapidity, $W_{\mu^{+}\mu^{-}}>10$ GeV for invariant mass, and $p_{t,\mu^-}>4$ GeV for transverse momentum. We have applied these restrictions in our calculation, and our results for the differential cross-section are shown as a function of invariant mass in Fig.\ref{cut_mu_2510}. In this figure, the area under the curve gives us the total cross-section value with all the restrictions mentioned above. 
In Table \ref{ATLAS-ALICE-total-cross}, we have tabulated our results for the muon pair production and compared them with the experimental result and STARLIGHT calculation. 
Our result is slightly lower than the experimental result and STARLIGHT calculation. Similar calculations have been done by using EPA method and the results can be found in references 
\cite{ MKGAS, AHMKGAS, CAVPGBDM, SRK, SKAHMBXFY, MKGRRWS, WZLRZTZXSY, AS}.

In conclusion, we have compared our calculations with the ATLAS and  ALICE experimental results. Although the measured physical quantities have been obtained with many restrictions, hence only the small portion of the produced leptons are observed, and they are not inconsistent with our calculations. This means that two-photon 
external field model is useful for predicting the electromagnetic lepton pair production mechanism. Most importantly, accurate calculations of the impact parameter dependent
cross-section help us in many other predictions such as multi lepton-pair production. Our calculations predict that probabilities of two or three pair production of electron and muon are quite high, and likely to be measured. There was an attempt to measure the  two-pair production cross sections \cite{vane, vane1} for $S + Au$ at 200 A GeV and $Pb +Au$ at 160 A GeV
collisions. Since the colliding energies are small compared to the LHC energies, multi-pair production was not experimentally detected. However, theoretical calculations of the multi-pair production cross sections become high with the increase of the colliding energies, therefore, it may help to measure the multi-pair productions of the leptons.
All these calculations and measurements help us for better understanding the QED in strong field.

Also, we have made some predictions for multi-pair production applying different experimental cuts on transverse momentum and rapidities of produced pairs. Our results are consistent in itself, and also consistent with other studies in the literature for multi-pair production(\cite{MKGAS}). Unlike the other studies in the literature, we also examined the double-muon pairs and  simultaneous production of electron-muon pairs in our study.
\begin{table} 
\centering
\caption{$\sigma_T$ values for the multi-pair production  from $Pb-Pb$ collision at $5.5$ TeV for different cuts on individual transverse momentums and rapidities of produced pairs. The first column shows the double electron-pair production, while the second column shows the double muon-pair production. The last group is for simultaneous production of electron and muon pairs.}
\scalebox{0.83}{%
\begin{tabular}{l l l l}
\hline 
\hline
\rule{0pt}{2.5ex} Cuts & $e^+ e^- e^+ e^-$  \quad \qquad & $\mu^+ \mu^- \mu^+ \mu^- $  \quad \qquad &  $e^+ e^- \mu^+ \mu^- $  \\[0.5ex] 
\hline 
\rule{0pt}{2.5ex} $p_{t_l}>0.2$ GeV \qquad\quad &  $70.1 \, \mu b$ & $64.9 \, \mu b$ & $66 \, \mu b $ \\ [0.5ex]
\rule{0pt}{2.5ex} $p_{t_l}>0.2$ GeV, $|y_l|<2.5$ \qquad\quad &  $13.44 \mu b $ & $ 9.3 \, \mu b $ & $9.38 \, \mu b $ \\ [0.5ex]
\rule{0pt}{2.5ex} $p_{t_l}>0.2$ GeV, $|y_l|<1$ \qquad\quad &  $0.538 \, \mu b$ & $0.453 \, \mu b $ & $0.476 \, \mu b $ \\ [0.5ex]
\\
\rule{0pt}{2.5ex} $p_{t_l}>0.3$ GeV, $|y_l|<4.9$ \qquad\quad &  $13.9 \, \mu b$ & $10.7 \, \mu b $ & $11.87 \, \mu b$ \\ [0.5ex]
\rule{0pt}{2.5ex} $p_{t_l}>0.3$ GeV, $|y_l|<2.5$ \quad\quad &  $2.62 \, \mu b$ & $ 2.32 \, \mu b$ & $2.47 \, \mu b $ \\ [0.5ex]
\\
\rule{0pt}{2.5ex} $p_{t_l}>0.5$ GeV, $|y_l|<4.9$ \qquad\quad &  $0.89 \,  \mu b$ & $ 1.12 \, \mu b $ & $0.97 \, \mu b $ \\ [0.5ex]
\rule{0pt}{2.5ex} $p_{t_l}>0.5$ GeV, $|y_l|<2.5$ \qquad\quad &  $0.285 \, \mu b$ & $0.34 \, \mu b $ & $0.288 \, \mu b $ \\ [0.5ex]
\\
\rule{0pt}{2.5ex} $p_{t_l}>1$ GeV \qquad\quad &  $44.4 \, nb$ & $ 47.5 \, nb$ & $45.5 \, nb$ \\ [0.5ex]
\rule{0pt}{2.5ex} $p_{t_l}>1$ GeV, $|y_l|<4.9$ \qquad\quad &  $40.7 \, nb$ & $43.6 \, nb $ & $41.9 \, nb $ \\ [0.5ex]
\rule{0pt}{2.5ex} $p_{t_l}>1$ GeV, $|y_l|<2.5$ \qquad\quad &  $13.2 \, nb$ & $18.1 \, nb $ & $14.7 \, nb $ \\ [0.5ex]
\rule{0pt}{2.5ex} $p_{t_l}>1$ GeV, $|y_l|<1$ \qquad\quad &  $0.75 \, nb$ & $0.98 \, nb $ & $0.81 \, nb $ \\ [0.5ex]
\hline
\hline
\end{tabular}}
\label{cuts_sigmaT_values}
\end{table}

\begin{acknowledgments}
This research is partially supported by the Istanbul Technical University.
\end{acknowledgments}


\begin{thebibliography}{99}	
	\bibitem{2}
	K. Kajantie, M. Kataja, L. McLerran, and P. V. Ruuskanen, Phys. Rev. D {\bf 34} (1986), 811.
	[K. Kajantie, J. Kapusta, L. McLerran, and A. Mekjian, Phys. Rev. D {\bf 34} (1986), 2746].
	\bibitem{11}
	C. R. Vane, S. Datz, P. F. Dittner, H. F. Krause, C. Bottcher, M. R. Strayer, R. Schuch, H. Gao,
	and R. Hutton, Phys. Rev. Lett. {\bf 69} (1992), 1911.
	\bibitem{bs} 
	C.~Bottcher and M.R.~Strayer, Phys. Rev. D {\bf 39}, 1330 (1989).
	\bibitem{bb} 
	C.A.~Bertulani and G.~Baur, Phys. Rep. {\bf 163}, 299 (1988).
	\bibitem{mcg} 
	M.~C.~G\"{u}\c{c}l\"{u}, J.C.~Wells, A.S.~Umar, M.R.~Strayer and D.J.~Ernst Phys. Rev. A {\bf 51}, 1836 (1995).
	\bibitem{mcgjl} 
	M.~C.~G\"{u}\c{c}l\"{u}, J.~Li, A.~S.~Umar, and D.~J.~Ernst, M.~R.~Strayer, Ann.~Phys. {\bf 272},7 (1999).     
	\bibitem{18}
	E. Fermi, Z. Phys. {\bf 29} (1924), 315. 
	\bibitem{ww}
	C.F. von Weizs\"{a}cker, Z. Phys, {\bf 88} (1934) 612. [E. J. Williams, Phys. Rev. {\bf 45} (1934), 730].
	\bibitem{ahtb}
	A. Alscher, K. Hencken, D. Trautmann, and G. Baur, Phys. Rev. A  {\bf 55} 396 (1997).
	\bibitem{bhatk}
	G. Baur, K. Hencken, A. Aste, D. Trautmann, S. R. Klein, Nucl. Phys. A  {\bf 729} 787 (2003).
	\bibitem{baur}
	C. A. Bertulani and G. Baur, Phys. Rep. {\bf 163} (1988), 299.
	\bibitem{ATLAS_mu}
	ATLAS collaboration, et al., ATLAS-CONF-2016-025 (2016).
	\bibitem{ALICE_el}
	E.,Abbas, $et \, al.$, Eur. Phys. Journal  C,{\bf 73(11)},2617 (2013).
	\bibitem{JadamALICE}
	J. Adam, $et \,al.$ (ALICE Collaboration), Phys. Rev. Lett. {bf 116}, 222301 (2016).
	\bibitem{MAatlas}
	M. Aboud, $et\, al.$ (ATLAS collaboration), Phys. Rev. Lett. {\bf 121}, 212301 (2018).
	\bibitem{MD}
	M. Dyndal, $et\, al.$ (ATLAS Collaboration), Nucl. Phys. A {\bf 967} (2017) 281.
	\bibitem{JadamSTAR}
	J. Adam, $et\, al.$ (STAR Collaboration), Phys. Rev. Lett. {\bf 121}, 132301 (2018).    
    \bibitem{LAstar}
   	L. Adamczyk, $et\, al.$ (STAR Collaboration), Phys. Lett. B 770 (2017) 451.
   	\bibitem{SAphenix}
   	S. Afanasiev, $et al.$ (PHENIX Collaboration), Phys. Lett. B 679 (2009) 321.
    \bibitem{Sengul}
    M.Y. Sengul, M. C. G\"{u}\c{c}l\"{u}, O. Mercan, N.G. Karakus, Eur. Phys. Journal  C {\bf 76}, 428 (2016).
    \bibitem{semi-analytic} 	
	M.~C.~G\"{u}\c{c}l\"{u}, Nuc. Phys. A {\bf 668}, 149 (2000).		
	\bibitem{erkan}
	E. Kurban, M.~C.~G\"{u}\c{c}l\"{u}, Phys. Rev. C {\bf 96(4)}, 044913(2017). 
    \bibitem{STARLIGHT}
	S. Klein and J. Nystrand, The STARLIGHT website, http://starlight.hepforge.org.
    \bibitem{MKGAS}
    M. K{\l}usek-Gawenda, A. Szczurek, Phys. Lett. B 763 (2016) 416.
    \bibitem{AHMKGAS}
    A. van Hameren, M. K{\l}usek-Gawenda, A. Szczurek, Phys. Lett. B 776 (2018) 84.
    \bibitem{CAVPGBDM}
    C. Azevedo, V.P. Gonçalves, B.D. Moreira, Eur. Phys. J. C (2019) 79:432.
    \bibitem{SRK}
    S.R. Klein, Phys. Rev. C {\bf 97}, 054903 (2018).
    \bibitem{SKAHMBXFY}
    S. Klein, A.H. Mueller, Bo-Wen Xiao, and F. Yuan, PRL {\bf 122}, 132301 (2019).
    \bibitem{MKGRRWS}   
    M. K{\l}usek-Gawenda, R. Rapp, W. Sch\"{a}fer, A. Szczurek, Phys. Lett. B 790 (2019) 339.
    \bibitem{WZLRZTZXSY}
    W. Zha, L. Ruan, Z. Tang, Z. Xu, S. Yang, Phys. Lett. B 781 (2018) 182.
    \bibitem{AS}
    A. Szczurek, Nucl. Phys. B 219-220 (2011) 17-24.
    \bibitem{vane}
    C.R. Vane, S. Datz, P.F. Dittner, H.F. Krause, R. Schuch, H. Gao, R. Hutton, Phys. Rev. A, 50 (1994), p. 2303.
    \bibitem{vane1}
    C.R. Vane, S. Datz, E.F. Deveney, et al. Phys. Rev. A  {\bf 56} 3682 (1997).
\end{thebibliography}

\end{document}